\documentclass{iopart}

\usepackage{iopams}
\usepackage{epsf}
\usepackage{cite}

\newcommand{\bd}{\begin{displaymath}}
\newcommand{\ed}{\end{displaymath}}
\newcommand{\be}{\begin{equation}}
\newcommand{\ee}{\end{equation}}
\newcommand{\ba}{\begin{eqnarray}}
\newcommand{\ea}{\end{eqnarray}}

\begin{document}

\title[On the influence of resonance photon scattering on atom
interference]
{On the influence of resonance photon scattering on atom interference}

\author{M Bo\v zi\'c$^1$, D Arsenovi\'c$^1$, A S Sanz$^2$ and
M Davidovi\'c$^3$}
\address{$^1$Institute of Physics, University of Belgrade, 11080 Belgrade,
Serbia}
\address{$^2$Instituto de F\'{\i}sica Fundamental - CSIC, Serrano 123,
28006 - Madrid, Spain}
\address{$^3$Faculty of Civil Engineering, University of Belgrade,
11000 Belgrade, Serbia}

\eads{\mailto{arsenovic@phy.bg.ac.yu},\mailto{bozic@phy.bg.ac.yu},
\mailto{asanz@imaff.cfmac.csic.es},\mailto{milena@grf.bg.ac.yu}}

\begin{abstract}
Here, the influence of resonance photon-atom scattering on the atom
interference pattern at the exit of a three-grating Mach-Zehnder
interferometer is studied.
It is assumed that the scattering process does not destroy the atomic
wave function describing the state of the atom before the scattering
process takes place, but only induces a certain shift and change of
its phase.
We find that the visibility of the interference strongly depends on the
statistical distribution of transferred momenta to the atom during the
photon-atom scattering event.
This also explains the experimentally observed (Chapman {\it et al}
1995 {\it Phys. Rev. Lett.} {\bf 75} 2783) dependence of the visibility
on the ratio $d_p/\lambda_i = y'_{12} (2\pi/kd\lambda_i)$, where
$y'_{12}$ is distance between the place where the scattering event
occurs and the first grating, $k$ is the wave number of the atomic
center-of-mass motion, $d$ is the grating constant and $\lambda_i$ is
the photon wavelength.
Furthermore, it is remarkable that photon-atom scattering events happen
experimentally within the Fresnel region, i.e.\ the near field region,
associated with the first grating, which should be taken into account
when drawing conclusions about the relevance of ``which-way''
information for the interference visibility.
\end{abstract}






\section{Introduction}
\label{sec1}

With the rise and advancement of neutron \cite{ref1} and atom
interferometry \cite{ref2,ref3}, it has become feasible the realization
of the well-known {\it gedanken} experiments devised by Einstein during
his famous discussions with Bohr \cite{ref4}, later also considered by
Feynman \cite{ref5}.
These discussions were focused on the understanding and interpretation
of the wave-particle duality and, therefore, the completeness of
Quantum Mechanics.
In particular, Bohr \cite{ref4} and Feynman \cite{ref5} argued that
wave and particle properties were complementary, i.e.\ they could not
be simultaneously observed experimentally.
On the other hand, aimed to disprove the concept of
{\it complementarity}, Einstein devised double-slit type experiments
\cite{ref4} where it should be possible to obtain ``which-way''
information without influencing the interference pattern.
Einstein's viewpoint based on the compatibility of the wave and
particle properties, i.e.\ that both are present simultaneously and,
therefore, can be observed in quantum interference experiments, was
supported by De Broglie \cite{ref6} and Bohm \cite{ref7}.
For these authors, the quantum system comprises both a wave {\it and}
a particle, the former guiding the motion (evolution) of the latter,
which leads to a hydrodynamic-like view of Quantum Mechanics
\cite{ref8,ref9}.

More recently, Rauch and Vigier have pointed out \cite{ref10} that de
Broglie's and Bohm's arguments were based on the so-called {\it einweg}
experiments, which should explicitly show that individual particles go
along one trajectory, but without evidencing which particular
trajectory.
Of course, the ``which-way'' argument implies the {\it einweg} one, but
not vice versa.
The difference between these two types of argument arises from the
different signatures of wave and particle properties invoked in them.
More specifically, for Bohr and Feynman the particle signature is the
``which-way'' information, while the wave signature is the visibility
or relative contrast of the interference pattern.
On the contrary, for de Broglie \cite{ref6}, Bohm \cite{ref7},
Philippidis {\it et al} \cite{ref11} or Sanz {\it et al} \cite{ref12,%
ref13,ref14}, the particle signature is the arrival of individual
quantum particles to a screen (array of detectors) and the time
evolution of the distribution of these arrivals \cite{ref15}; the wave
signature associated with each quantum particle is the visibility of
the interference pattern together with the fact that it comes from the
accumulation of arrivals of a large number (theoretically, an infinite
number) of atoms, photons, electrons, etc.

To perform experimentally Feynman's {\it gedanken} experiment, Chapman
{\it et al} \cite{ref16} scattered single photons from Na atoms within
a three-grating Mach-Zehnder atom interferometer.
By measuring the transmission of atoms through the third grating,
the influence of photon scattering processes (which take place at a
distance $y'_{12}$ from the first grating) on the visibility of the
atom interference pattern was investigated.
These results have intensified a controversial discussion on the
wave-particle duality issue.
At the time when the experiment was carried out, this controversy
evolved towards a discussion around the question: Is complementarity
more fundamental than the uncertainty principle? \cite{ref17,ref18,%
ref19,ref20}
This discussion continued \cite{ref21,ref22,ref23,ref24,ref25} with
the aim to determine the cause of the visibility decrease: Does the
visibility decrease arise (a) from a random momentum transfer between
the atom and the photon or (b) from the correlations between the
``which-way'' detector and the atomic motion?
More recently, the statement (b) has been reformulated as \cite{ref3}:
Is the visibility decrease (decoherence) the result of entanglement
between a quantum system (the atom) and an environment (the emitted
photon, which carries information about the atom's path).

Previously, we explained \cite{ref26} the experiment carried out
by Chapman {\it et al} using the solution of the time-dependent
Schr\"odinger equation for an atom interacting with a photon and
the gratings in a three-grating Mach-Zehnder interferometer.
In our explanation, wave and particle properties were compatible, since
in our opinion both are present and play a role. We derived an analytic
expression for the visibility dependence on the ratio $d_p/\lambda_i =
y'_{12} (2\pi/kd\lambda_i)$, where $k$ is the wave number of the atomic
center-of-mass motion, $d$ is the grating constant and $\lambda_i$ is
the photon wavelength.
This theoretical result was in fairly good agreement with the
visibility measured in the experiment \cite{ref16}.
The distribution of transferred momentum during the photon scattering
process leads to the visibility decrease as $d_p/\lambda_i$ approaches
0.5 as well as several subsequent revivals with decreasing maxima.
Here, we provide additional arguments which support our conclusions of
reference \cite{ref26}.
In particular, we study the visibility dependence on the features
of the atom selection at the exit of the interferometer, before the
detection takes place.
We also show that in the experiment the photon-atom interaction takes
place within the Fresnel region, i.e.\ the near field region, associated
with the first grating, something that has to be taken into account
within any dynamical (i.e.\ time-dependent) description of the
experiment.


\section{Wave function of an atom after interacting with a grating and
a photon}
\label{sec2}

Consider an initial stationary atomic monochromatic wave, which spreads
along the $y$-axis and is incident to a one-dimensional grating
parallel to the $x$-axis at $y=0$,
\begin{equation}
 \Psi(x,y,t) = e^{-i\omega t} \psi^i (x,y)
  = B^i e^{-i\omega t} e^{iky} , \qquad y < 0 ,
 \label{eq1}
\end{equation}
with $B^i$ being a constant.
After interacting with the grating (i.e.\ after getting diffracted),
this incident wave transforms into
\begin{equation}
 \Psi(x,y,t) = e^{-i\omega t} \psi (x,y) ,
 \label{eq2}
\end{equation}
where
\begin{equation}
 \psi(x,y) = \frac{e^{iky}}{\sqrt{2\pi}}
  \int_{-\infty}^\infty dk_x c(k_x) e^{ik_x x} e^{-ik_x^2 y/2k} ,
 \label{eq3}
\end{equation}
satisfies the Helmholtz equation \cite{ref27}.
If the grating is completely transparent inside the slits (the union of
slit areas is denoted by $A$) and completely absorbing outside them,
$c(k_x)$ can be expressed \cite{ref27} as
\begin{eqnarray}
 c(k_x) & = & \frac{1}{\sqrt{2\pi}}
  \int_{-\infty}^\infty dx' \psi(x',0^+) e^{-ik_x x'} \nonumber \\
  & = & \frac{1}{\sqrt{2\pi}}
  \int_A d_x \psi^i (x',0^-) e^{-ik_x x'} ,
 \label{eq4}
\end{eqnarray}
where $\psi^i (x',0^-)$ and $\psi (x',0^+)$ denote the wave function
just before and just after the first grating, respectively.
The solution of the Helmholtz equation, $\psi(x,y)$, given by
(\ref{eq3}), is equivalent to the Fresnel-Kirchhoff solution
\begin{equation}
 \psi(x,y) = \sqrt{\frac{k}{2\pi y}} \ \! e^{-i\pi/4} e^{iky}
  \int_{-\infty}^\infty dx' \psi(x',0^+) e^{ik(x - x')^2/2y} .
 \label{eq5}
\end{equation}

The photon-atom scattering event induced by the laser light at a
distance $y'_{12}$ from the first grating leads to a change of the
atomic transverse momentum, $\Delta k_x$, and, therefore, to a shift
of the wave function in the momentum representation.
Hence, after an atom absorbs and re-emits again a photon somewhere
along the $x$ axis at a time $t'_{12}$ and a distance $y'_{12} =
v t'_{12} = (\hbar k/m) t'_{12}$ from the first grating, the atomic
wave function takes the form \cite{ref26}
\begin{eqnarray}
 \psi_{\Delta k_x}(x,y) & = & \frac{e^{iky}}{\sqrt{2\pi}} \ \!
  e^{i\Delta k_x (x + \Delta x_0) - i \Delta k_x^2 y/k} \nonumber \\
 & & \times \int_{-\infty}^\infty dk'_x c(k'_x) e^{-i{k'}_x^2 y/2k}
   e^{ik'_x (x + \Delta x_0 - \Delta k_x y/k)} ,
 \label{eq6}
\end{eqnarray}
where
\begin{equation}
 \Delta x_0 = \frac{\Delta k_x \hbar t'_{12}}{m}
  = \frac{\Delta k_x y'_{12}}{k} .
 \label{eq7}
\end{equation}
The wave function (\ref{eq6}), evaluated at the distance $y_{12}$,
which separates the second from the first grating, was used
\cite{ref26} as a wave incident onto the second grating.
Then, the wave function propagating towards the third grating was
determined using (\ref{eq5}), where $\psi(x',0^+)$ consists of the
parts of the incident wave which are transmitted through the slits
of the second grating.

\begin{figure}
 \begin{center}
 \epsfxsize=8cm {\epsfbox{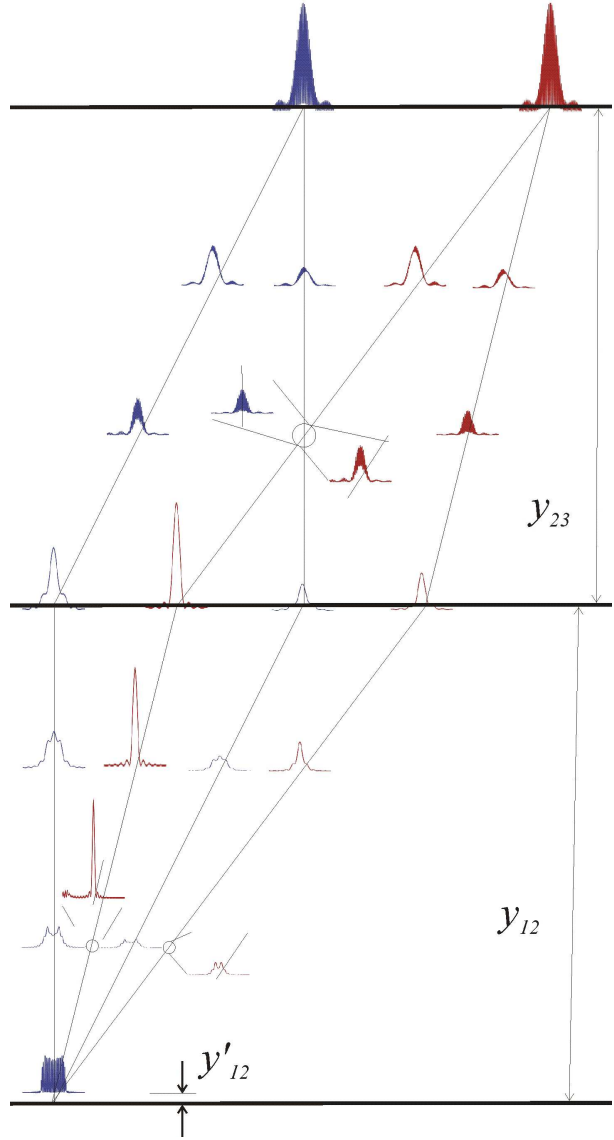}}
 \caption{\label{fig1}
  Atom probability density within a three-grating Mach-Zehnder
  interferometer at several distances from the first grating for two
  values of the transferred momentum: $\Delta k_x = 0$ (blue) and
  $\Delta k_x = 1.5k_i$ (red).
  As in the experiment, the incoming particles are Na atoms and the
  parameters used are: $v_{\rm Na} = 1400$~m/s,
  $k = m_{\rm Na} v_{\rm Na}/\hbar = 5.09 \times 10^{11}$~m$^{-1}$,
  $k_i = 2\pi/(589\ {\rm nm}) = 1.07 \times 10^7$~m$^{-1}$, $y_{12}
  = y_{23} = 65$~cm, $d = 200$~nm, slit width $\delta = 100$~nm and
  number of slits illuminated by the incident atomic plane wave $n=24$.
  The distance $y'_{12} = 2.863$~mm corresponds to $d_p/\lambda_i =
  0.3$~mm, which lies within the near field region, this being evident
  from the fact that the Talbot distance is $L_t = 2d^2/\lambda =
  6.484$~mm.}
 \end{center}
\end{figure}

This is the way how the evolution of the initial plane wave (\ref{eq1})
was determined.
After interacting with the first grating, a photon at the distance
$y'_{12}$ from this slit, and a second grating, the resulting wave
function evolves freely again up to the third grating.
In order to illustrate this evolution, in figure~\ref{fig1} we have
plotted the atom probability density within the interferometer at
several distances from the first grating.
Two values for the transferred momentum are considered: $\Delta k_x=0$
(blue) and $\Delta k_x=1.5k_i$ (red).
Far from a grating, the straight lines represent the paths along which
the maxima of the probability density move; near the grating, these
straight lines would just be the prolongation of the paths.
Note that in the near field associated with a grating, the wave
function has a very complex form and the lines do not exactly represent
the paths of the atoms.
Within this region, according to a Bohmian picture \cite{ref13,ref14},
there are many paths which, as the distance from a grating increases,
converge towards three main paths (only two are plotted in
figure~\ref{fig1}).
The extension of the near field is of the order of $10L_T$, where
$L_t = 2d^2/\lambda$ is the so-called {\it Talbot distance}
\cite{ref14}.

\section{Dependence of visibility on the distribution of transferred
momentum}
\label{sec3}

In the experiment \cite{ref16}, the number of Na atoms transmitted
through the third grating is counted for the laser off and laser on
varying the distance $y'_{12}$ as well as considering different values
of the shift $\Delta x_3$ of the third grating along the $x$-axis.
In a first round of measurements, all transmitted atoms were collected
(counted) without carrying out any selection.
Thus, the resulting interference curve can be directly associated with
the distribution of transferred momentum, which is given by the
Mandel-Wolf expression \cite{ref28,ref29}
\begin{equation}
 P_{MW} (\Delta k_x) = \frac{3}{8k_i}
  \left[ 1 + \left( 1 - \frac{\Delta k_x}{k_i}\right)^2 \right] .
 \label{eq8}
\end{equation}
Then, in subsequent measurements, specific subsets of transmitted atoms
were counted after being selected using certain slits positioned after
the third grating.
As explained in \cite{ref16}, each selection is equivalent to a
particular distribution of transferred momentum during the photon-atom
scattering process.
The experimental data show that the interference pattern visibility
strongly depends on the ratio $d_p/\lambda_i$ as well as on the
probability distribution for the transferred momentum, $P(\Delta k_x)$.

In order to explain and interpret these experimental data within our
approach \cite{ref26}, it is necessary to study the function
\begin{equation}
 T(y'_{12}, \Delta x_3) = \int_0^{2k_i} d(\Delta k_x) P(\Delta k_x)
  \tilde{T}(y'_{12}, \Delta k_x, \Delta x_3) ,
 \label{eq9}
\end{equation}
where
\begin{equation}
 \tilde{T}(y'_{12}, \Delta k_x, \Delta x_3) =
  \int_{slits} \left\arrowvert \psi_{\Delta k_x} (x,y=y_{12}+y_{23})
    \right\arrowvert^2 dx .
 \label{eq10}
\end{equation}
The latter function is proportional to the number of atoms transmitted
through the third grating that undergone a change of momentum $\Delta
k_x$ during the photon-atom scattering process behind the first
grating.
By numerical integration, it was shown \cite{ref26} that (\ref{eq10})
has the general form
\begin{equation}
 \tilde{T}(y'_{12}, \Delta k_x, \Delta x_3) =
  a + b \cos (2\pi\Delta x_3/d + d_p \Delta k_x) ,
 \label{eq11}
\end{equation}
where $a$ and $b$ are constants, and
\begin{equation}
 d_p = (2\pi/k d) y'_{12} .
 \label{eq12}
\end{equation}
For $y'_{12}$ far from a grating, $d_p$ denotes the separation between
two paths associated with the zeroth and first order interference
maxima; near the grating, it refers to the distance between the
prolongations of these paths.
Based on these results, we may assume that for certain classes of
distribution functions $P (\Delta k_x)$, $T(y'_{12},\Delta x_3)$ has
the general form
\begin{equation}
 T(y'_{12}, \Delta x_3) = a + b V \cos (2\pi\Delta x_3/d + \varphi) ,
 \label{eq13}
\end{equation}
where the visibility $V$ and phase-shift $\varphi$ are both functions
of the ratio $d_p/\lambda_i$, and their particular form is determined
by $P(\Delta k_x)$, which is assumed to be normalized on the interval
$[0,2k_i]$.

In order to understand the features of $V(d_p/\lambda_i)$ and $\varphi
(d_p/\lambda_i)$, consider, for instance, the uniform distribution
$P(\Delta k_x) = 1/2k_i$.
The evaluation of the integral of the first term in $T(y'_{12},\Delta
x_3)$ is trivial, and for the second term we have
\begin{eqnarray}
 & & \int_0^{2k_i} \frac{1}{2k_i} \ \! d(\Delta k_x) b
  \cos (2\pi\Delta x_3/d + d_p \Delta k_x) \nonumber \\
 & & \qquad \quad = \frac{b}{k_i d_p} \ \! \sin (d_p k_i)
   \cos (2\pi\Delta x_3/d + d_p k_i) .
 \label{eq14}
\end{eqnarray}
From this result, we reach
\begin{equation}
 T(y'_{12},\Delta x_3) = a +
   \frac{b}{k_i d_p} \ \! \sin (d_p k_i)
   \cos (2\pi\Delta x_3/d + d_p k_i) .
 \label{eq15}
\end{equation}
and, therefore,
\begin{equation}
 V = \frac{1}{k_i d_p} \ \! \sin (d_p k_i) \qquad {\rm and} \qquad
  \varphi = d_p k_i .
 \label{eq16}
\end{equation}
As can be seen, $V$ vanishes for $d_p k_i = n\pi$, which is easy to
explain as follows.
The integrand in (\ref{eq14}) is a periodic function of $\Delta k_x$,
with period $2\pi/d_p$.
Hence, for
\begin{equation}
 \frac{2k_i}{2\pi/d_p} = \frac{2d_p}{\lambda_i} = n , \qquad
  {\rm i.e.\ for} \qquad \frac{d_p}{\lambda_i} = \frac{n}{2} ,
 \label{eq17}
\end{equation}
the integration is performed over an integer number of periods of a
simple periodic function, the result being zero.

If we now consider the distribution $P_{MW}(\Delta k_x)$, described by
(\ref{eq8}), we obtain the transmission function \cite{ref26}
\begin{equation}
 T(y'_{12},\Delta x_3) = a + b V \cos (2\pi\Delta x_3/d + d_p k_i) ,
 \label{eq18}
\end{equation}
where the visibility reads as
\begin{equation}
 V = \frac{3}{4\pi} \frac{\lambda_i}{d_p} \left[
  \left( 1 - \frac{1}{(2\pi)^2} \frac{\lambda_i^2}{d_p^2} \right)
  \sin \left( \frac{2\pi d_p}{\lambda_i} \right) +
  \frac{1}{2\pi} \frac{\lambda_i}{d_p}
  \cos \left( \frac{2\pi d_p}{\lambda_i} \right) \right] .
 \label{eq19}
\end{equation}
The zeros of (\ref{eq19}) are very close to those of the visibility
obtained from the constant distribution above (see figure~2 in
reference \cite{ref26}).

Finally, if we consider the distribution
\begin{equation}
 P_I (\Delta k_x) = \gamma e^{-(\Delta k_x/Nk_i)^2} ,
 \label{eq20}
\end{equation}
where $\gamma = 2/Nk_i\sqrt{\pi}$, we obtain
\begin{equation}
 V = \frac{|{\rm erf} (2/N - i\alpha) + {\rm erf} (i\alpha)|}
     {{\rm erf} (2/N)}\ e^{-\alpha^2/4} ,
 \label{eq21}
\end{equation}
\begin{equation}
 \varphi = \frac{1}{2i}\ \ln \left[
  \frac{{\rm erf}(2/N - i\alpha) + {\rm erf}(i\alpha)}
    {{\rm erf}(2/N + i\alpha) + {\rm erf}(-i\alpha)} \right] ,
 \label{eq22}
\end{equation}
where $\alpha = N k_i d_p$ and ${\rm erf}$ is the error function,
${\rm erf}(\alpha) \equiv \frac{2}{\sqrt{\pi}}\ \int_0^\alpha e^{-t^2}
dt$.
The visibility given by (\ref{eq21}) does not have zeros; for any
$d_p/\lambda_i$, this function is always above the values of the
visibility arising from the previous two distributions.
This is in agreement with the experimental data displayed in figure~2
of reference \cite{ref16}.
From this fact, Chapmann {\it et al} concluded \cite{ref16} that
{\it coherence was not really destroyed, but only entangled with
the final state of the reservoir} (photons).

As can be seen, our results agree with the first part of the
conclusions of Chapmann {\it at al} \cite{ref16}.
In evaluating the visibility, we have assumed that the scattering
process does not destroy the atomic wave function (i.e.\ coherence),
which describes the state of the atom before the scattering process,
but only induces a certain shift and change of its phase.
The visibility dependence on $d_p/\lambda_i$ is thus a consequence of
the statistical distribution of transferred momenta to the atom during
photon-atom scattering process.
Through the atom selection, which is equivalent to varying the
distribution of transferred momenta, the visibility can change
substantially.
Within our approach, we have not invoked
entanglement with the reservoir states; rather, we have determined
and used the evolution of an atomic wave function before and after
photon-atom scattering events.

Finally, we would like to stress that experimentally photon-atom
scattering events take place at distances from a grating within
the Fresnel region.
This can be easily seen by means of the relation
\begin{equation}
 y'_{12} = \frac{kd}{2\pi} \ \! d_p
  = \frac{d_p}{\lambda_i} \frac{kd}{k_i}
  = \frac{d_p}{\lambda_i} \frac{L_T}{2} \frac{\lambda_i}{d} .
 \label{eq23}
\end{equation}
In the experiment, the ratio $d_p/\lambda_i$ goes from 0 to 2.
By the values of
the other parameters given in the caption of figure~\ref{fig1}, it
follows that $y'_{12} \in [0, 19.09]$~mm, with $L_T = 6.48$~mm.


\section{Conclusions}
 \label{sec4}

To explain atom interference experiments with presence of photon-atom
scattering processes, in our opinion it is necessary to use the atom
wave function as well as to take into account its particle properties
(i.e.\ the change of momentum during photon-atom scattering events).
The experimentally established visibility dependence on $d_p/\lambda_i$
was previously explained \cite{ref26} by considering a random change of
the atomic transverse momentum induced by the scattering with photons.

The experimental regain of visibility induced by selecting a subset of
atoms from the set of all those transmitted through the third grating
is explained by studying the visibility dependence on the probability
distribution of transferred momenta.
This atom selection does not provide any information about the place
along the $x$-axis where the photon scatters from the atom.
Consequently, it is not necessary to attribute the decrease and
disappearance of visibility as $d_p/\lambda_i$ increases to an increase
of an observer's (potential) knowledge about the atomic path behind the
first grating.

From our description, we also find that photon-atom scattering
processes happen within the Fresnel region, where the atomic wave
function has a very complex form (see figure~\ref{fig1}).
But, as has been shown within the context of the Talbot effect
\cite{ref14}, the topology of the trajectories also becomes very
complex within this region.
Therefore, not two but many atomic paths exist near the grating, where
the photon hits the atom.
As one moves further away from the grating (towards the Fraunhofer
region), those numerous trajectories basically group along three main
paths.
Here, we have considered two of them for a chosen value of the
transferred momentum to explain the experiment.

The agreement between our theoretical expressions for the visibility
and the experimental curves thus supports, in our opinion, the views
of Einstein, de Broglie, Bohm and others, i.e.\ that individual
micro-objects can be simultaneously described by a wave and a particle.
In particular, here we have stressed that such an agreement has been
obtained by using both the space and momentum atom distributions as
quantum objects.
In this way, our results also agree with and support the more general
conclusions due to Ballentine \cite{Ballentine} and Khrennikov
\cite{Khrennikov1,Khrennikov2}.
According to these authors, the question of complementarity versus
compatibility of wave and particle properties is tightly connected to
the problem of the existence of a joint distribution of two dynamical
variables associated with two non-commutative operators, as well as
to the question of whether simultaneous measurements of two
non-commutative operators is possible.
In our theoretical approach, we have not considered a joint
distribution of coordinates and momenta, but we used both
distributions to explain the experimental results.

To recapitulate, our results show that different outcomes of the
experiment should not be associated with the presence of an external
observer, but with the sensitivity of the experiment which is being
carried out or, in other words, the particular measurement performed.
More specifically, with our analysis we have shown here that the
visibility depends on the experiment itself and can be nicely explained
taking into account all the elements present in such an experiment.

The importance of context, i.e. the concrete specification of the
experimental setup in the analysis of interference phenomena is
reconsidered by Ballentine \cite{Ballentine} and Khrennikov
\cite{Khrennikov1,Khrennikov2} from a general point of view.
According to the latter author, the main structures of the Quantum
Theory are already present in a latent form within the classical
Kolmogorov probability model.
In this regard, a very interesting question arises, namely to compare
our findings and method with such an approach.
At present, this idea is being considered as a direction for future
research.


\ack

M.\ Bo\v zi\'c, D.\ Arsenovi\'c and M.\ Davidovi\'c acknowledge support
from the Ministry of Science of Serbia under Project ``Quantum and
Optical Interferometry'', N 141003.
A.\ S.\ Sanz acknowledges support from the Ministerio de Ciencia e
Innovaci\'on (Spain) under Project FIS2007-62006 and the Consejo
Superior de Investigaciones Cient\'{\i}ficas for a JAE-Doc Contract.


\end{document}